\documentclass[twocolumn,showpacs,amsmath,amssymb,superscriptaddress,aps,prc]{revtex4-1}
\usepackage[dvips]{graphicx}
\usepackage{epsfig}
\usepackage{amsmath}
\usepackage[dvips]{color}
\newcommand{\beq}{\begin{eqnarray}}
\newcommand{\eeq}{\end{eqnarray}}
\newcommand{\li}{\mbox{$^{11}$Li}}
\newcommand{\be}{\mbox{$^{14}$Be}}
\newcommand{\bo}{\mbox{$^{15}$B}}
\newcommand{\n}{\mbox{$^{17}$N}}
\newcommand{\neon}{\mbox{$^{17}$Ne}}
\newcommand{\neonq}{\mbox{$^{15}$Ne}}
\newcommand{\linn}{\mbox{$^{9}$Li+n+n}}
\newcommand{\bnn}{\mbox{$^{13}$B+n+n}}
\newcommand{\nnn}{\mbox{$^{15}$N+n+n}}
\newcommand{\opp}{\mbox{$^{15}$O+p+p}}
\newcommand{\oopp}{\mbox{$^{13}$O+p+p}}
\newcommand{\benn}{\mbox{$^{12}$Be+n+n}}
\newcommand{\kmax}{\mbox{$K_{\rm max}$}}
\newcommand{\rmax}{\mbox{$R_{\rm max}$}}

\begin{document}
	\title{Microscopic three-cluster study of light exotic nuclei}
	\author{P. Descouvemont}
	\email{pdesc@ulb.ac.be}
\affiliation{Physique Nucl\'eaire Th\'eorique et Physique Math\'ematique, C.P. 229,
	Universit\'e Libre de Bruxelles (ULB), B 1050 Brussels, Belgium}
\date{\today}
\begin{abstract}
I develop a microscopic three-cluster model for exotic light nuclei. I use the hyperspherical formalism,
associated with the Generator Coordinate Method. This model is well adapted to halo nuclei, since the 
long-range part of the radial wave functions is accurately reproduced.
The core wave functions are described in the shell model,
including excited states. This technique provides large bases, expressed in terms of projected Slater determinants.
Matrix elements involve seven-dimension integrals, and therefore require long calculation times.
I apply the model to $\li$, $\be$, $\bo$, and $\n$ described by two neutrons surrounding a
$^9$Li, $^{12}$Be, $^{13}$B and $^{15}$N core, respectively. 
The $\neon$ (as $\opp$) and $\neonq$ (as $\oopp$) mirror nuclei are briefly discussed.
I present the spectra and some spectroscopic properties,
such as r.m.s.\ radii or E2 transition probabilities. I also analyze the importance of core excitations.
\end{abstract}
\maketitle

\section{Introduction}
Exotic nuclei represent one of the major interests in modern nuclear physics \cite{TSK13}.  These nuclei 
are located close to the driplines (neutron or proton) and, owing to the low binding energy, present 
specific properties, such as an anomalously large radius or modifications of the shell structure.  
Recent development of experimental facilities have provided a large number of new data, which need 
to be understood by theory.

The main property of exotic nuclei being their low binding energy, the relative wave functions extend to 
large distances.  Theoretical models therefore need to reproduce this property.  A widely used 
approach is the hyperspherical method \cite{Li95}, where the Jacobi coordinates are replaced by a 
set of angles, and by a single length, referred to as the hyperradius. The three-body equation is 
then replaced by a set of coupled differential equations depending on the hyperradius.  
The hyperspherical formalism can be extended to systems involving more than three particles \cite{GKV11}.  
Many works in atomic and in nuclear physics have been performed within this method.

In nuclear physics, most applications are carried out in non-microscopic models.  In other words, 
the nucleus is seen as a three-body system, with a structureless core, and two external nucleons.  
Typical applications are the $^{6}$He  and $\li$ nuclei, modeled by $\alpha$+n+n and $\linn$ 
three-body structures.  This approach therefore relies on nucleon+nucleon and nucleon+core 
potentials.  It simulates the Pauli principle by an appropriate choice of these interactions.
Core excitations are in general neglected.

This non-microscopic model can be extended to microscopic theories, where the system is described by 
a $A$-body Hamiltonian.  The core nucleus is defined in the shell model, and a full antisymmetrization 
is taken into account.  The use of the hyperspherical formalism guarantees the correct long-range 
behaviour of the wave function.  The model only relies on a nucleon-nucleon interaction, and on the 
description of the core wave functions.  In contrast with non-microscopic models, it allows 
to include core excitations without any further parameter.

Microscopic cluster models are used in nuclear physics for many years (see reviews in Refs.\ \cite{DD12,HIK12}) 
but most applications deal with the two-cluster variant, much easier than multicluster approaches.  The 
application of the hyperspherical formalism to microscopic cluster theories is recent.  The first works 
were  focused on $^6$He \cite{FKK96,KD04}  which involves the $\alpha$ particle as a core.  The $\alpha$ particle 
can be accurately described by a $(0s)^4$ shell model configuration which makes the calculations relatively easy.  
The model was extended to three-body continuum states \cite{DD09}, and microscopic $^6$He wave functions have 
been used in CDCC calculations of elastic scattering on heavy targets \cite{De16b}.

The main limitation of microscopic three-cluster models in hyperspherical coordinates is that the matrix 
elements involve the numerical calculation of (many) seven-dimension integrals \cite{KD04}.  Consequently, 
applications are essentially limited to light systems, most of them involving an $\alpha$ core.  In addition to the  $^6$He 
system mentioned above, a recent application focuses on  $^8$Li, described by an $\alpha+^3$H+n three-cluster 
configuration \cite{DP19}.

The aim of the present work is to go beyond this limitation.  The development of the computing technology, 
and in particular of the parallelization possibilities permits to consider heavier systems, involving 
$p$-shell nuclei as the core.  I analyze four nuclei ($\li$, $\be$, $\bo$, $\n$) which are modelled by a 
$p$-shell core and two surrounding neutrons.  Compared to previous applications, the present 
calculations face two difficulties: (1) the presence of $p$-shell orbitals, and (2) the need of 
several Slater determinants for the core.  

The paper is organized as follows.  In Sec.\ \ref{sec2}, I present a general overview of the 
microscopic three-cluster model.  In 
Sec.\ \ref{sec3}, I apply the method to various exotic nuclei.  Concluding remarks are presented 
in Sec.\ \ref{sec4}.

\section{Microscopic three-cluster model in hyperspherical coordinates}
\label{sec2}
\subsection{General overview}
My main goal is to solve a $A$-body problem, where the Hamiltonian is given by
\beq
H=\sum_{i=1}^A t_i +\sum_{i<j=1}^A v_{ij}.
\label{eq1}
\eeq
In this definition, $t_i$ is the kinetic energy of nucleon $i$, and $v_{ij}$ a nucleon-nucleon interaction 
which contains nuclear and Coulomb components.  Recent {\sl ab initio} models (see, for example, 
Ref.\ \cite{BNV13}) are developed to find exact solutions of the Schr\"odinger equation associated 
with (\ref{eq1}).  These models, however, are in general not well adapted to halo nuclei, where the 
long-range part of the wave functions plays a crucial role.  My approach, in contrast, is based 
on the cluster approximation \cite{WT77,DD12}. 

The total wave function, solution of the Schr\"odinger equation
\beq
(H-E)\Psi=0,
\label{eq1b}
\eeq
is written schematically as
\beq
\Psi={\cal A}\phi_1\phi_2\phi_3 G,
\label{eq2}
\eeq
where $\phi_i$ are the internal wave functions of the clusters (in practice, they are defined in the 
shell model), and $G$ is a radial function, depending on the relative coordinates between the clusters.  
The antisymmetrization operator ${\cal A}$ ensures that the wave function is completely antisymmetric.  

The cluster approximation permits a strong simplification of the calculations, in comparison with 
{\sl ab initio} models.  It is also well adapted to halo nuclei or, more generally, to states presenting
a strong cluster structure.  Owing to this approximation, however,
effective nucleon-nucleon interactions $v_{ij}$, 
such as the Volkov \cite{Vo65} or the Minnesota \cite{TLT77} force, must be used.  
These forces simulate missing effects, such as the tensor component, by an effective central interaction. Ideally, three-body forces should be introduced, and developments have been
achieved in that direction for realistic forces (see Ref.\ \cite{MMM13} for a review). For effective interactions, however, even if some work
has been done \cite{KA04,It16}, the use of three-body forces is essentially limited to well bound nuclei,
such as $^{12}$C or $^{16}$O. In most cluster calculations, three-body forces are therefore neglected.  

Several variants of microscopic cluster models exist.  Of course, the simplest version is a two-cluster 
model, which is used since many years \cite{Ho77}.  Three-cluster models have been developed 
in various directions: frozen triangular configurations \cite{De95}, 2+1 configurations essentially 
aimed for nucleus-nucleus scattering such as $^7$Be+p \cite{De04}, and, more recently, genuine three-body 
models using the hyperspherical formalism \cite{KD04}.  The present work is based on the 
hyperspherical approach, which is described in more detail in the next subsections.

\subsection{Core wave functions}
Let me consider Eq.\ (\ref{eq2}) where the internal wave functions $\phi_i$ are associated with three clusters 
with nucleon numbers ($A_1,A_2,A_3$) and charge numbers ($Z_1,Z_2,Z_3$).  I assume that clusters 2 and 3 
are $s$-shell nuclei and, more specifically in the present work, that they are neutrons. I also
 assume that the oscillator
parameter $b$ is common to the three clusters. Until now, the use of the hyperspherical formalism 
in microscopic models was limited to an $\alpha$ core.  This is well adapted, for example, to 
$^6$He\cite{KD04}, $^6$Li \cite{KD04}, or $^8$B \cite{DP19}.  The main reason for this limitation is that, 
as it will be discussed later, matrix elements involve seven-dimensional integrals.  

This limitation, however, restricts the applications to a few light nuclei.  
In the present work, I go beyond this limitation, and extends the model to $p$-wave orbitals.  
The consequences in terms of computer times are twofold: ($i$) the matrix elements involve $p$-shell 
orbitals which has a strong impact on the matrix elements of the nucleon-nucleon interaction (quadruple sums over 
the individual orbitals); ($ii$) $p$-shell nuclei such as $^9$Li or $^{12}$Be involve several 
Slater determinants (up to 90 for $^9$Li), whereas the $\alpha$ particle is described by a single Slater determinant.

I remind here the main properties.  Let me consider a Slater determinant $\bar{\Phi}_i$ involving 
$A_1$ single-particle orbitals
\begin{align}
&\varphi_{n_x n_y n_z m_s m_t}(\pmb{r})= \nonumber \\
&\hspace{1cm}\varphi_{n_x}(x) \varphi_{n_y}(y) \varphi_{n_z}(z) \vert \frac{1}{2} m_s \rangle 
\vert \frac{1}{2} m_t \rangle ,
\label{eq3}
\end{align}
where $\varphi_{n}(s)$ are harmonic oscillator functions and where $m_s,m_t=\pm 1/2$ are associated 
with the spin and the isospin, respectively.  The first step is to define the list of $N_S$ Slater determinants 
consistent with the Pauli principle.  Assuming that the $s$-shell is filled, the number of Slater 
determinants for $p$-shell nuclei is $N_S=C_6^{Z_1-2}\times C_6^{N_1-2} $, where $C_i^j$ is the 
number of combinations of $j$ elements among $i$ elements.  
For $^9$Li, $^{12}$Be, $^{13}$B and $^{15}$N, I have $N_S=90,15,20,6$, respectively.   
The $N_S$ Slater determinants $\bar{\Phi}_i$ must be projected on the various angular
momenta $I,L,S,T$ from a diagonalization of the $\pmb{I}^2,\pmb{L}^2,\pmb{S}^2$ and $\pmb{T}^2$ operators
($\pmb{I}=\pmb{L}+\pmb{S}$).  
This procedure provides basis functions with good quantum numbers as
\beq
\Phi^{I\nu }_{1,LST}=\sum_{i=1}^{N_S} d^{I\nu LST}_i \, \bar{\Phi}_i.
\label{eq4}
\eeq
Finally, basis functions (\ref{eq4}) are used to diagonalize the Hamiltonian (\ref{eq1}) for cluster 1, 
and I get the core wave functions as
\beq
\Phi^{I\nu}_{1}=\sum_{LST} D^{I\nu}_{LST} \Phi^{I\nu}_{1,LST}.
\label{eq5}
\eeq
This technique corresponds to a standard shell-model approach, and can be extended to higher shells
(the summation may involve an additional quantum number, associated with the degeneracy).  
The specificity of the cluster model is that (\ref{eq5}) is only the very first step of the calculations.  
These internal wave functions must then be introduced in the three-body wave functions (\ref{eq2}).

\subsection{Three-cluster wave functions}
\begin{figure}[htb]
	\begin{center}
		\epsfig{file=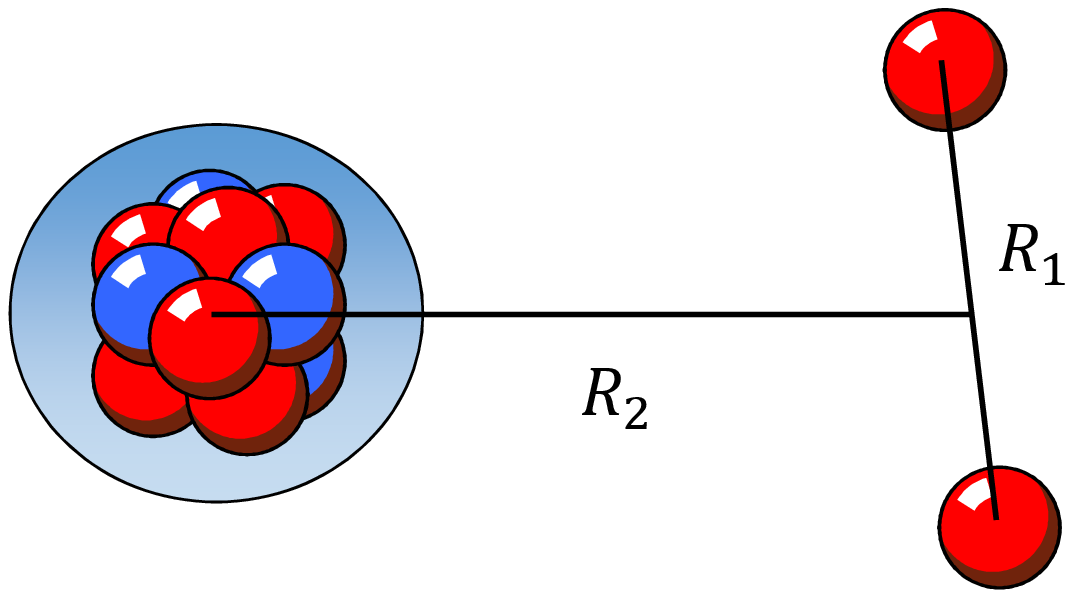,width=6.0cm}
		\caption{Three-cluster configuration using the generator coordinates $\pmb{R}_1$ and $\pmb{R}_2$
			[see Eq.\ (\ref{eq6})].}
		\label{fig_conf}
	\end{center}
\end{figure}

Let me come back to the three-cluster wave functions (\ref{eq2}).  In the Generator Coordinate Method 
(GCM, see Refs.\ \cite{Ho77,DD12}), the clusters are located at three points, as represented in 
Fig.\ \ref{fig_conf}.  The generator coordinates $\pmb{R}_1$ is associated with the external clusters
(neutrons in the present applications) and $\pmb{R}_2$ with the relative motion between the core and
the c.m.\ of clusters 2 and 3. They are variational parameters, and 
are not associated with the nucleon coordinates (in other words, the antisymmetrization operator
does not act on $\pmb{R}_1$ and $\pmb{R}_2$).  A GCM basis state is defined as
\begin{align}
&\Phi^{\nu_1\nu_2\nu_3}_{c_1}(\pmb{R}_1,\pmb{R}_2)={\cal A} 
\Phi^{\nu_1}_{c_1}\bigl(-\frac{A_{23}}{A}\pmb{R}_2\bigr)\nonumber \\
&\hspace{0.5cm}\times \Phi^{\nu_2}_{2}\bigl(\frac{A_{23}}{A}\pmb{R}_2+\frac{A_{1}}{A_{12}}\pmb{R}_1\bigr) 
 \Phi^{\nu_3}_{3}\bigl(\frac{A_{23}}{A}\pmb{R}_2-\frac{A_{1}}{A_{12}}\pmb{R}_1\bigr),
\label{eq6}
\end{align}
where $\Phi^{\nu_1}_{c_1}$ are the core wave functions (\ref{eq5}), and $\Phi^{\nu_2}_{2}$ and where
$\Phi^{\nu_3}_{3}$ correspond to the external clusters.  In this equation,  
$A_{12}=A_1+A_2 $ and $A_{23}=A_2+A_3$.  If the internal wave functions are defined in the harmonic oscillator 
model with a common oscillator parameter $b$, these basis functions are Slater determinants, and 
the c.m. motion is factorized exactly.  For the sake of clarity, I do not write the spins $I_i$ of the clusters. 

A first angular-momentum coupling is performed on the spins $I_2$ and $I_3$ of the external clusters with 
\begin{align}
&\Phi^{I_{23}I\nu}_{c_1}(\pmb{R}_1,\pmb{R}_2)=\sum_{\nu_1\nu_2\nu_3}
\langle I_2 \, \nu_2 \, I_3\,  \nu_3 \vert I_{23}\,  \nu_2 + \nu_3 \rangle \nonumber \\
&\times \langle I_1\,  \nu_1 I_{23}\,  \nu_2 + \nu_3 \vert I \, \nu \rangle
\Phi^{\nu_1\nu_2\nu_3}_{c_1}(\pmb{R}_1,\pmb{R}_2).
\label{eq7}
\end{align}
Then I introduce the hyperspherical formalism, well known in non-microscopic models \cite{ZDF93}, and 
extended more recently to microscopic theories \cite{KD04,KRV08}. I define scaled Jacobi 
coordinates as
\begin{align}
 &\pmb{X}=\sqrt{\frac{A_2 A_3}{A_{23}}} \pmb{R}_1, \nonumber \\
 &\pmb{Y}=\sqrt{\frac{A_1 A_{23}}{A}} \pmb{R}_2,
\label{eq8}
\end{align}
which provide the hyperradius $R$ and hyperangle $\alpha$ from
\begin{align}
& R=\sqrt{ \pmb{X}^2+\pmb{Y}^2}, \nonumber \\
& \tan \alpha=Y/X.
\label{eq9}
\end{align}
Both coordinates are complemented by the angles associated with $\pmb{X}$ and $\pmb{Y}$ as 
\begin{align}
\Omega_5=(\alpha,\Omega_X,\Omega_Y).
\label{eq10}
\end{align}

From the basis functions (\ref{eq7}), I project on the total angular momentum $J$ and parity $\pi$.  
This is achieved by a double projection on the orbital momenta $\ell_x$ and $\ell_y$; a projected basis
function is written as
\begin{align}
&\tilde{\Phi}^{JM\pi}_{c_1 I_{23} I \ell_x \ell_y L}(R_1,R_2)=\sum_{\nu M_L}
	\langle I \nu L M_L \vert JM \rangle \nonumber \\
&\hspace{0.2cm}\times \int d\Omega_X d\Omega_Y
\bigl[Y_{\ell_x}^{\ast}(\Omega_X)\otimes Y_{\ell_y}^{\ast}(\Omega_Y)\bigr]^{LM_L}
 \Phi^{I_{23}I\nu}_{c_1}(\pmb{R}_1,\pmb{R}_2),
\label{eq11}
\end{align}
which represents a four-dimension integral.  This double-projection technique has been used in three-body 
models aimed at studying nucleus-nucleus scattering such as $^7$Be+p \cite{DB94} for example.  In 
the hyperspherical formalism, one introduces the hypermoment $K$, which is a generalization of the 
angular momentum in three-body systems.  An hyperspherical basis function reads
\begin{align}
\Phi^{JM\pi}_{\gamma K}(R)=&\int d\alpha \cos ^2 \alpha \sin ^2 \alpha  
\, \varphi_K^{\ell_x \ell_y}(\alpha) \nonumber \\
&\times \tilde{\Phi}^{JM\pi}_{\gamma}(R\sin  \alpha, R \cos \alpha),
\label{eq12}
\end{align}
where index $\gamma$ stands for $\gamma=(c_1 I_{23} I \ell_x \ell_y L)$. In this equation, function
$\varphi_K^{\ell_x \ell_y}(\alpha)$ is defined by
\begin{align}
\varphi_K^{\ell_x \ell_y}(\alpha)=&{\cal N}_K^{\ell_x \ell_y} (\cos \alpha_R)^{\ell_y} (\sin \alpha_R)^{\ell_x}
\nonumber \\
&\times P_n^{\ell_y+1/2,\ell_x+1/2}(\cos 2\alpha),
\label{eq13}
\end{align}
$P_n^{a,b}(x)$ being a Jacobi polynomial. The normalization coefficient ${\cal N}_K^{\ell_x \ell_y}$ is given
by
\begin{align}
&{\cal N}_K^{\ell_x \ell_y} =
\bigg[ \frac{2 n! (K+2) (n+\ell_x+\ell_y+1)!}
{\Gamma(n+\ell_x+\frac{3}{2}) \Gamma(n+\ell_y+\frac{3}{2})} \bigg]^{\frac{1}{2}}, 
\label{eq14} 
\end{align}
where $n=(K-\ell_x-\ell_y)/2$ is a positive integer.  The main advantage of the hyperspherical 
formalism is to reduce the number of degrees of freedom to one.  The configuration space is therefore 
spanned by a single generator coordinate $R$.  This approach is also well adapted to three-body continuum 
states \cite{DD09}.  Of course, matrix elements between basis states (\ref{eq12}) are more demanding 
in terms of computer times, since they involve seven-dimensional integrals (see next subsection) which
must be performed numerically.  
 
The total wave function of the system is given by a superposition of basis functions (\ref{eq12}) as
\begin{align}
\Psi^{JM\pi}=\sum_{\gamma K}\sum_{i=1}^N f^{J\pi}_{\gamma K i} \, \Phi^{JM\pi}_{\gamma K}(R_i),
\label{eq15} 
\end{align}
where $N$ is the number of generator coordinates (typically $N\approx 10$).  In this expansion, the 
summation over the hypermoment $K$ is limited to a maximum value $\kmax$.  As usual
in hyperspherical models, $\kmax$ must be large 
enough to ensure the convergence of the physical quantities (energies, radii, etc.).  
Coefficients $f^{J\pi}_{\gamma K i}$ represent the generator function, and are obtained from a diagonalization of the Hamiltonian kernel.

\subsection{Matrix elements}
The main part of the calculation concerns matrix elements between projected basis functions (\ref{eq12}).  
The Hamiltonian kernel reads
\begin{align}
&H^{J\pi}_{\gamma K,\gamma' K'}(R,R')=
\langle \Phi^{J\pi}_{\gamma K}(R)\vert H \vert  \Phi^{J\pi}_{\gamma' K'}(R') \rangle \nonumber \\
&=\int  \cos ^2 \alpha \sin ^2 \alpha \cos ^2 \alpha' \sin ^2 \alpha'
\, \varphi_K^{\ell_x \ell_y}(\alpha) \, \varphi_{K'}^{\ell_x' \ell_y'}(\alpha') \nonumber \\
&\times \tilde{H}^J_{\gamma,\gamma'}(R\sin \alpha,R\cos \alpha;R'\sin \alpha',R'\cos \alpha') d\alpha\, d\alpha',
\label{eq16} 
\end{align}
where
\begin{align}
\tilde{H}^{J\pi}_{\gamma,\gamma'}(R_1,R_2;R'_1,R'_2)=
\langle 
\tilde{\Phi}^{J\pi}_{\gamma}(R_1,R_2)\vert H \vert \tilde{\Phi}^{J\pi}_{\gamma'}(R'_1,R'_2)\rangle,
\label{eq17} 
\end{align}
with $\tilde{\Phi}^{J\pi}_{\gamma}(R_1,R_2)$ given by Eq.\ (\ref{eq11}).  The projection over the hypermoment 
therefore represents a double integration over $\alpha$ and $\alpha'$.  This quadrature is performed 
numerically. From Eq.\ (\ref{eq11}), the matrix elements 
(\ref{eq17}) involve eight angular integrals.  In fact, owing to the rotation invariance, three angles 
can be fixed, which leads to five-dimensional integrals.  The calculation provides
\begin{widetext}
\begin{align}
&\langle 
\tilde{\Phi}^{J\pi}_{\gamma}(R_1,R_2)\vert H \vert \tilde{\Phi}^{J\pi}_{\gamma'}(R'_1,R'_2)\rangle=\frac{8\pi^2}{2J+1}
\sum_{M_L,M'_L,\nu,\nu'}\langle I \, \nu \, L \, M_L\vert J \, \nu+M_L\rangle
\langle I' \, \nu' \, L' \, M'_L\vert J \, \nu'+M'_L\rangle \nonumber \\
&\times \int  
\bigl[Y^{\ast}_{\ell_x}(\Omega_{X'}) \otimes Y^{\ast}_{\ell_y}(0,0)\bigr]^{L M_L}
\bigl[Y_{\ell_{x'}}(\Omega_{X'}) \otimes Y_{\ell_{y'}}(\theta_{Y'},0)\bigr]^{L' M'_L}\nonumber \\
&\times \langle \Phi_{c_1} ^{I_{23} I \nu}(\pmb{R}_1,\pmb{R}_2) \vert H \vert 
\Phi_{c'_1} ^{I'_{23} I' \nu'}(\pmb{R}'_1,\pmb{R}'_2) \rangle\,
d\Omega_X d\Omega_{X'} \sin \theta_{Y'} d\theta_{Y'},
\label{eq18} 
\end{align}
\end{widetext}
where $\pmb{R}_2$ is along the $z$ axis and $\pmb{R}'_2$ is in the $xz$ plane.  This expression is also 
valid for other rotation-invariant operators. A generalization to operators, such as the electromagnetic
operators, is straightforward.

The calculation of the matrix elements (\ref{eq17}) is therefore performed in several steps:
\begin{enumerate}
\item  Matrix elements between Slater determinants ${\cal A} \bar{\Phi}_i \Phi^{\nu_2}_{2} \Phi^{\nu_3}_{3}$
 are first computed.  One-body operators 
involve double sums over the individual orbitals, whereas two-body operators involve quadruple sums.  
When dealing with three $s$ clusters, the quadruple sums contain $3^4$ terms, but this amounts to $6^4$ when 
one cluster belongs to the $p$ shell.  Compared to previous works on $^6$He or $^{12}$C this property makes 
the calculations 16 times longer.

\item  Matrix elements between wave functions (\ref{eq6}) are then constructed with the transformation 
coefficients (\ref{eq5}).  The number of three-cluster Slater determinants is given by $N_S (2I_2+1)
(2I_3+1)$.  For $^6$He, this number is 4, since the core is an $\alpha$ 
particle.  For $\li$, in contrast, the core involves 90 functions, and the total number of Slater determinants (\ref{eq6}) is therefore 360.

\item  The next step is to compute the projected matrix elements (\ref{eq18}) which involve five 
numerical quadratures (typically $\sim 16-20$ points are used for each angle).

\item  Finally the projection over the hypermomentum is performed with (\ref{eq16}).  For a given set of 
generator coordinates $(R,R')$, all integrals are evaluated simultaneously.
\end{enumerate}

This process must be repeated for all sets of $(R,R')$ values.  For systems such as $\li$ which involve
many Slater determinants, the full 
calculation is extremely time consuming.  This can be achieved with an optimization of the codes, 
and using modern computing facilities.  In practice, a parallelization is performed over the hyperangles 
$\alpha$ and $\alpha'$ in (\ref{eq16}).  Let me also mention that the total number of basis functions 
(i.e.\ the size of the matrices) can be as large as 20000.  This raises precision issues since the 
basis is (highly) non orthogonal.

\section{Application to light exotic nuclei}
\label{sec3}
\subsection{Conditions of the calculations}

The calculations are performed with the Volkov V2 nucleon-nucleon interaction \cite{Vo65}.  For the 
oscillator parameter, I use $b=1.60$ fm, a standard value for $p$-shell nuclei.  
The optimal $b$ value stems from a compromise: to reproduce the core radius, and to minimize the binding energy of the core. Small variations of $b$ can be compensated by a slight readjustment of the nucleon-nucleon interaction.
The first step is 
to determine the core wave functions (\ref{eq5}), as mentioned in Sec.\ \ref{sec2}.B.  Including 
all $p$-shell configurations provides $(90,15,20,6)$ Slater determinants $\bar{\Phi}_i$ [see Eq.\ (\ref{eq4})] 
for $^9$Li, $^{12}$Be, $^{13}$B and $^{15}$N, respectively.  The possible quantum numbers 
$(I,L,S,T)$ are discussed in the Appendix. For $^9$Li which has a
$I=3/2^-$ ground state, I limit the states to $I=1/2,3/2$ and to
$T=3/2$ to keep the size of the basis within acceptable values.

Values of the generator coordinate $R$ are selected from 1.5 to 15 fm with a step of 1.5 fm.  For 
the maximum hypermomentum, I use $\kmax=16$.  These conditions are sufficient to ensure the convergence 
of the energies and r.m.s. radii.

The central V2 interaction is complemented by a zero-range spin-orbit force with amplitude $S_0$ \cite{KD04}.  
This amplitude is fixed to $40$ MeV.fm$^5$, except for $\li$, where I use $S_0=50$ MeV.fm$^5$.  
The Volkov potential contains the admixture parameter $M$, whose standard value is $M=0.6$.  This value, 
however, can be slightly modified without changing the fundamental properties of the interaction.  
The aim is to reproduce exactly the ground-state energies.  For weakly bound nuclei, the properties 
are sensitive to the long-range part of the wave function, and therefore to the binding energies.  
With $M=0.7750,0.6119,0.5945$ and $0.6177$, I reproduce the experimental two-neutron separation 
energies $S_{2n}$ of $\li$, $\be$, $\bo$, and of $\n$, respectively (0.370 MeV,
1.27 MeV, 3.75 MeV and 8.37 MeV \cite{AKM17}).

\begin{figure}[htb]
	\begin{center}
		\epsfig{file=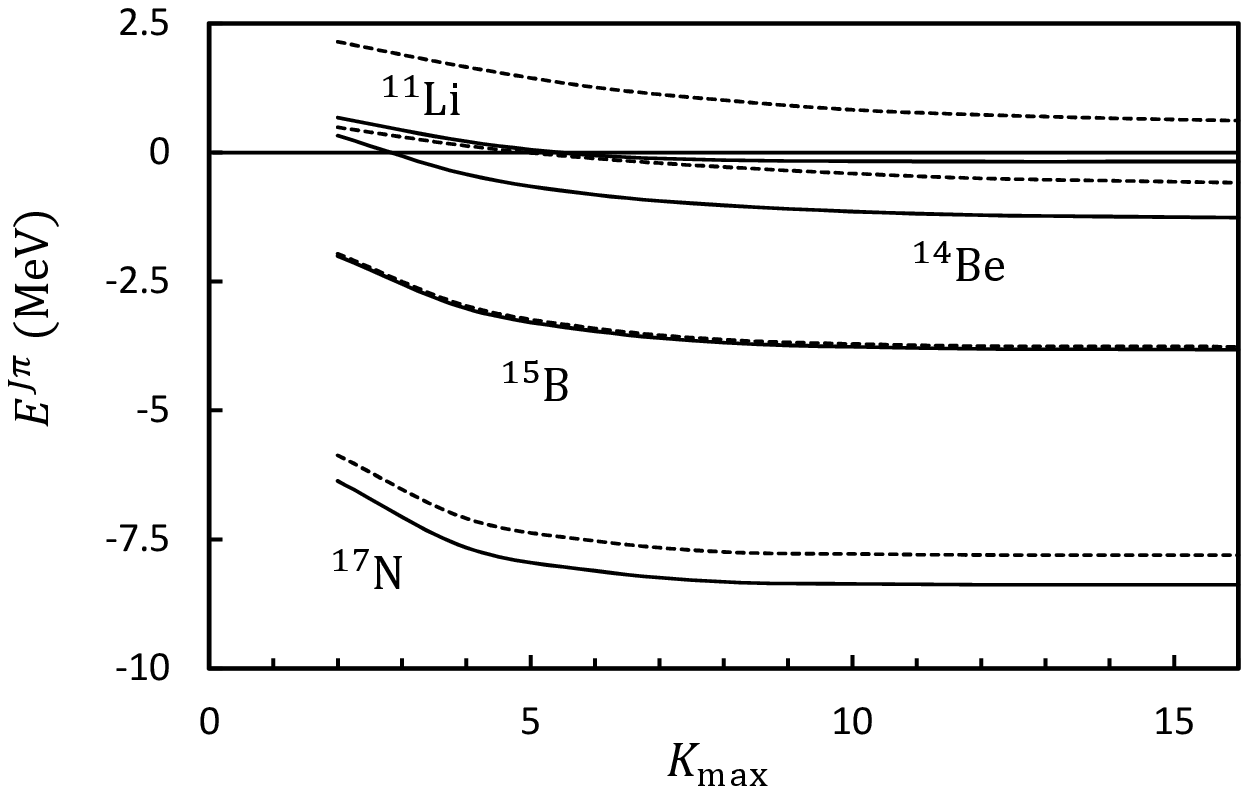,width=8.6cm}
		\caption{Ground-state energies (with respect to the core+n+n threshold) as a function of the
	maximum hypermomentum $\kmax$. The dotted lines are obtained by neglecting core excitations.}
		\label{fig_kmax}
	\end{center}
\end{figure}

In Fig.\ \ref{fig_kmax}, I present the convergence of the ground-state energies with the maximum 
hypermomentum $\kmax$.  In all cases, $\kmax=16$ guarantees a good convergence.  In this general
overview, I also evaluate the importance of core excitations. In Fig.\ \ref{fig_kmax}, the dotted lines
represent the energies obtained by neglecting core excitations. This is done in Eq.\ (\ref{eq19}) by keeping
only $\gamma$ values corresponding to the ground state of the core. In $\be$ on in $\bo$, this effect is of the order of 0.6 MeV. It is quite small in $\n$: core excitations are negligible in the ground state. For $\li$,
the ground state is unbound if core excitations are neglected. The energy difference is of
the order of 0.9 MeV.

I discuss now the 
properties of each nucleus by increasing mass, except for $\li$ which I present as last application since, 
by far, it is the most complicated.

\subsection{$\be$ as a $\benn$ system}
Before considering a full diagonalization of the basis, I first display the energy curves, where a 
single value of the generator coordinate $R$ is considered.  The energy curves $E^{J\pi}_i(R)$ are therefore 
obtained from the eigenvalue problem

\begin{align}
& \sum_{K' \gamma'} \bigl[ H^{J\pi}_{\gamma K,\gamma' K'}(R,R)-
E^{J\pi}_i(R) N^{J\pi}_{\gamma K,\gamma' K'}(R,R)\bigr] \nonumber \\
& \hspace{1cm} \times f^{J\pi}_{\gamma' K'}=0,
\label{eq19}
\end{align}
where $N^{J\pi}_{\gamma K,\gamma' K'}(R,R')$ is the overlap kernel. 
They provide a useful overview of the system.  From the attractive or repulsive character, one can predict 
the existence of bound states (or of narrow resonances).  In all cases, the three-body threshold is subtracted.

\begin{figure}[htb]
	\begin{center}
		\epsfig{file=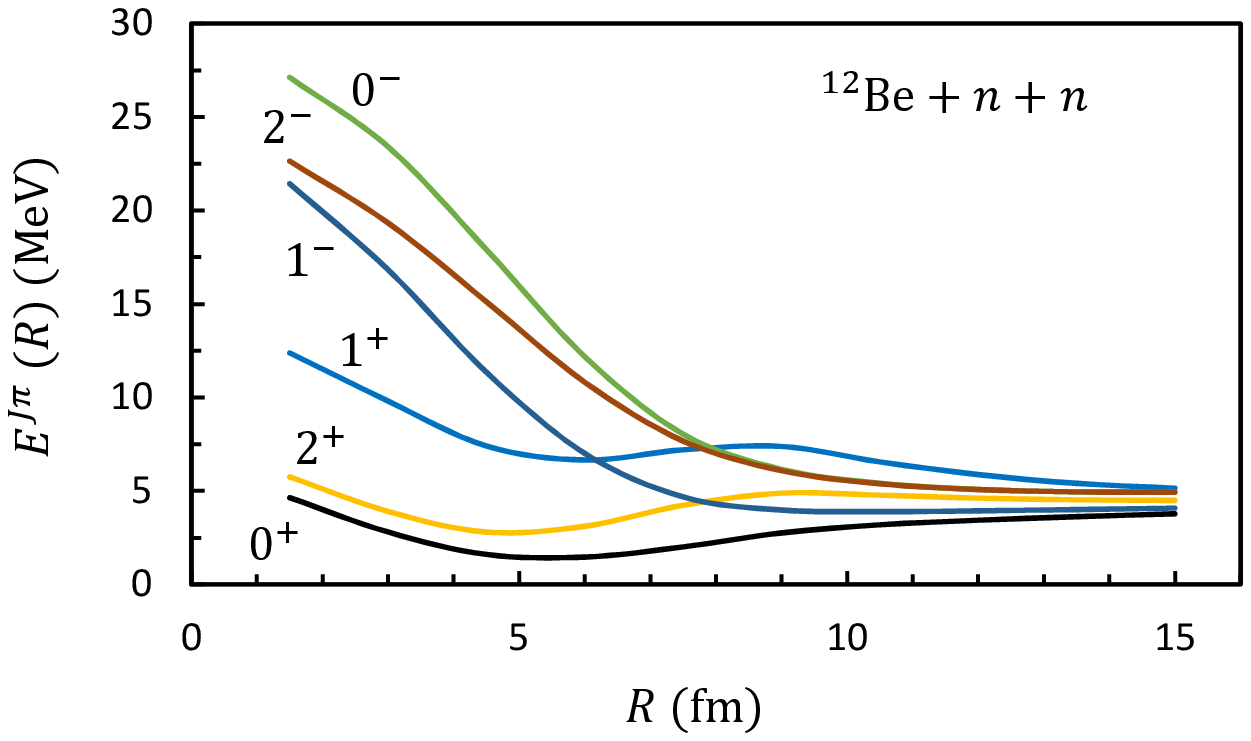,width=8.6cm}
		\caption{$\benn$ energy curves (\ref{eq19}) for different $J\pi$-values.}
		\label{fig_be14_ce}
	\end{center}
\end{figure}

The $\benn$ energy curves are displayed in Fig.\ \ref{fig_be14_ce}.  In positive parity, there is a minimum 
for $J=0^+$ and $J=2^+ $, which correspond to the $\be$ ground state and to the $2^+$ resonance, respectively.  
In negative parity, the $0^-$ and $2^-$ curves are repulsive.  There is a shallow minimum for $J=1^-$ which 
might be associated with a broad resonance.  A deeper analysis of such resonances, however, would require 
a specific formalism for continuum states \cite{DD09}, and is beyond the scope of the present work.

\begin{figure}[htb]
	\begin{center}
		\epsfig{file=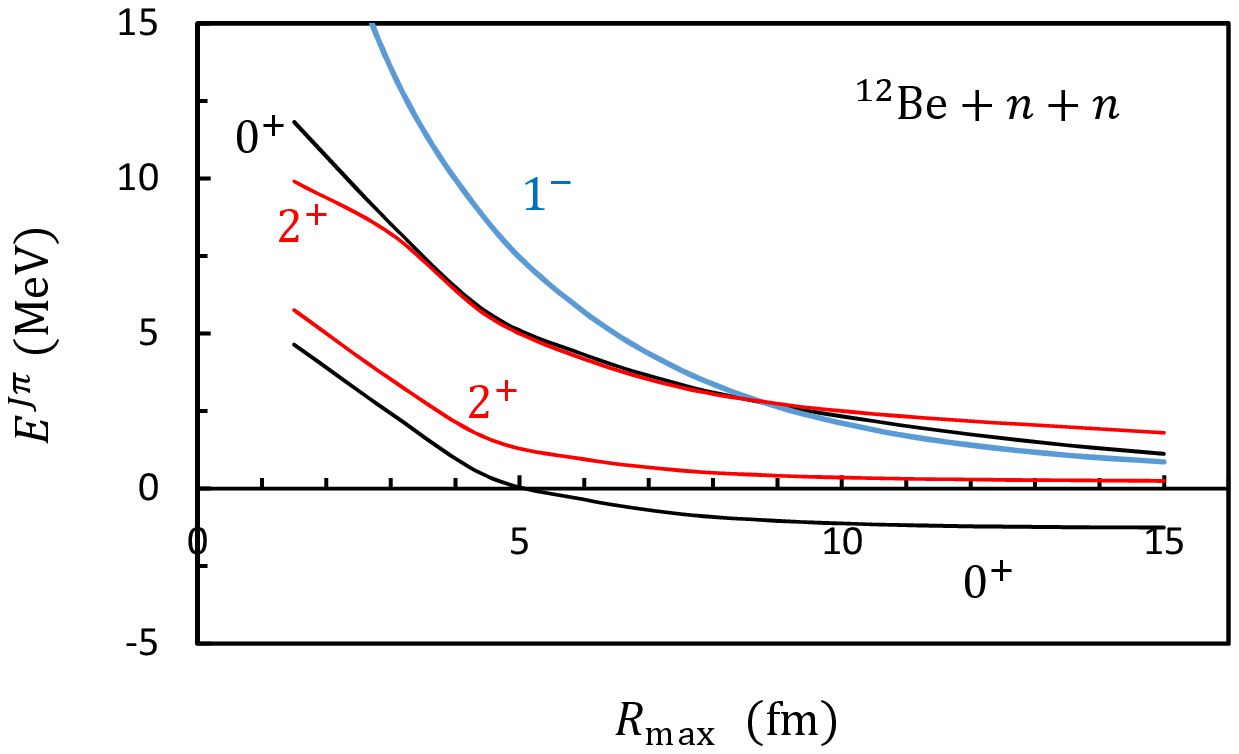,width=8.6cm}
		\caption{Convergence of $\be$ energies with respect to the maximum generator coordinate $\rmax$.}
		\label{fig_be14_rmax}
	\end{center}
\end{figure}

An interesting characteristic of the system is provided by the energy convergence as a function of 
the maximum $R$ value, which I denote as $\rmax$.  In Fig.\ \ref{fig_be14_rmax}, I show $\be$ energies 
obtained by increasing the $N$ value in Eq.\ (\ref{eq15}) or, in other words, by increasing $\rmax$.  
Two different behaviours can be clearly observed.  The $0^+$ and $2^+$ energies are almost stable 
above $\rmax \approx 10$ fm, and present a plateau.  The $2^+$ energy is 0.25 MeV, in nice agreement 
with experiment ($0.28 \pm 0.01$ MeV \cite{SNK07}).  This result emphasizes the importance of a 
microscopic theory; a non-microscopic three-body model, based on $^{12}$Be+n and on $n+n$ 
phenomenological interactions, does not reproduce the $0^+$ and $2^+$ energies simultaneously \cite{DTB06}.
The $^{12}$Be(g.s.)+n+n component is 87\% in the ground state, and 67\% in the $2^+$ resonance. This means that
core excitations play a role, and may explain why non-microscopic models, which ignore core excitations, cannot
predict the $2^+$ energy accurately.
The other curves in Fig.\ \ref{fig_be14_rmax} do not present a plateau, which means that no further 
narrow resonance can be expected. In particular, the $1^-$ curve is typical of a continuum state where,
according to the variational principle, the minimum energy is zero.
 
\begin{figure}[htb]
	\begin{center}
		\epsfig{file=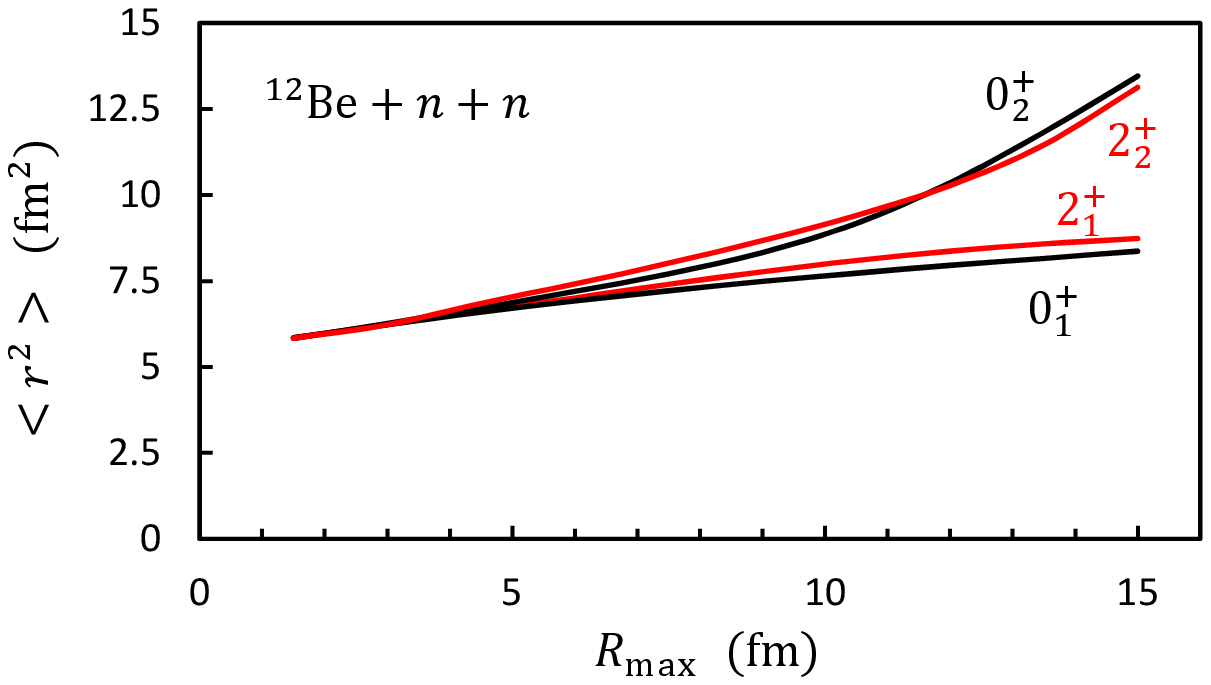,width=8.6cm}
		\caption{Convergence of the $\be$ squared radius as a function of the maximum generator coordinate
			$\rmax$. The two first $0^+$ and $2^+$ eigenvalues are shown.}
		\label{fig_be14_r2}
	\end{center}
\end{figure}

Another useful method to distinguish between narrow and broad resonances 
is to analyze the convergence of the r.m.s. radius $\langle r^2 \rangle$ as a function of $\rmax$.  
This is displayed in Fig.\ \ref{fig_be14_r2}.  The $0^+$ and $2^+$ low-lying states reach convergence 
near $\rmax\approx 12$ fm.  In contrast, the second eigenvalues present a diverging behaviour at 
large $\rmax$.  This is expected since, strictly speaking, matrix elements involving continuum states diverge.  
Although I cannot specifically address continuum states, this technique allows a clear distinction 
between narrow and broad resonances. It is consistent with the energy convergence shown in Fig.\ \ref{fig_be14_rmax},
and confirms that r.m.s.\ radii in the continuum should be considered very carefully.

The ground-state radii are presented in Table \ref{table2}, and compared with experiment.  The neutron radius is 
in nice agreement with experiment.  The experimental proton radius, however, is significantly larger than
the predicted value.  This is difficult to understand from a $\benn$ model, where the $\be$ proton radius
should be close to the core proton radius.
Experimental values are partly model dependent, and a reanalysis of the $\be$ charge 
radius would be welcome.

\begin{table}[htb]
	\caption{Proton, neutron and matter radii (in fm).
		\label{table2}}
	\begin{ruledtabular}
		\begin{tabular}{lcccccc}
	    & \multicolumn{2}{c}{$\sqrt{<r^2>_p}$}  & \multicolumn{2}{c}{$\sqrt{<r^2>_n}$}& \multicolumn{2}{c}{$\sqrt{<r^2>_m}$}   \\
	    \hline
	&Th.  & Exp.  & Th.  & Exp. & Th.  & Exp.   \\
	\hline
$\be$ & 2.10 & $3.00(36)$\tablenotemark[1]& 3.24 & $3.22(39)$\tablenotemark[1]& 2.96 & $3.16(38)$\tablenotemark[1]  \\	
$\bo$ & 2.23 & $2.56(8)$\tablenotemark[2]& 2.82 & & 2.64 & $2.75(6)$\tablenotemark[2] \\	
$\n$ & 2.32 & & 2.61 & & 2.49 & $2.48(5)$\tablenotemark[3]  \\	
$\neon$ & 2.69 & $3.04(2)$\tablenotemark[4]& 2.32 & & 2.54 & $2.75(7)$\tablenotemark[4]  \\	
$\li$ & 1.97 & $2.467(37)$\tablenotemark[5]& 3.12 & $3.36(24)$& 2.85 & $3.27(24)$\tablenotemark[6]  \\	
\end{tabular}
\end{ruledtabular}
\tablenotetext[1]{Ref.\ \cite{TKY88}.}
\tablenotetext[2]{Ref.\ \cite{EKH14}.}
\tablenotetext[3]{Ref.\ \cite{OKS94}.}
\tablenotetext[4]{Ref.\ \cite{GNA08}.}
\tablenotetext[5]{Ref.\ \cite{SNE06}.}
\tablenotetext[6]{Ref.\ \cite{THH85}.}
\end{table}

Notice that the radii are slightly sensitive to the oscillator parameter $b$. From the total matter radius
$\sqrt{<r^2>_m}$, one can define the expectation value of the hyperradius $<R^2>$ from
\beq
A<r^2_m>=A_1<r_1^2>+<R^2>,
\eeq
where $\sqrt{<r_1^2>}$ is the core radius. In the shell model, this quantity is proportional to $b$ and
takes the values $<r_1^2>=49b^2/24$ for $^{12}$Be, $<r_1^2>=27b^2/13$ for $^{13}$B, $<r_1^2>=32b^2/15$ for $^{15}$N
and $^{15}$O, and  $<r_1^2>=17b^2/9$ for $^{9}$Li. In contrast $<R^2>$, which is associated with the external
neutrons, weakly depends on the oscillator parameter. For compact states, $<R^2>$ is small and the matter 
radius approximately varies linearly with $b$. For halo states, however, $<R^2>$ is the dominant term,
and the matter radius is almost insensitive to the oscillator parameter.

\subsection{$\bo$ as a $\bnn$ system}
The $\bo$ ground state is known to be bound by 3.77 MeV.  Although the spin assignment is no definite, 
there are strong indications for a spin $J=3/2^-$.  Two excited states at $E_x=1.336$
MeV and $E_x=2.743$ MeV have been reported in Ref.\ \cite{KEB05}.  On the theoretical side, 
shell-model \cite{WB92} and Antisymmetrized Molecular Dynamics (AMD, see Ref.\ \cite{KH95}) calculations 
have been performed.  Non-microscopic calculations are unavailable until now, essentially due to the lack
of reliable $^{13}$B+n potentials.

\begin{figure}[htb]
	\begin{center}
		\epsfig{file=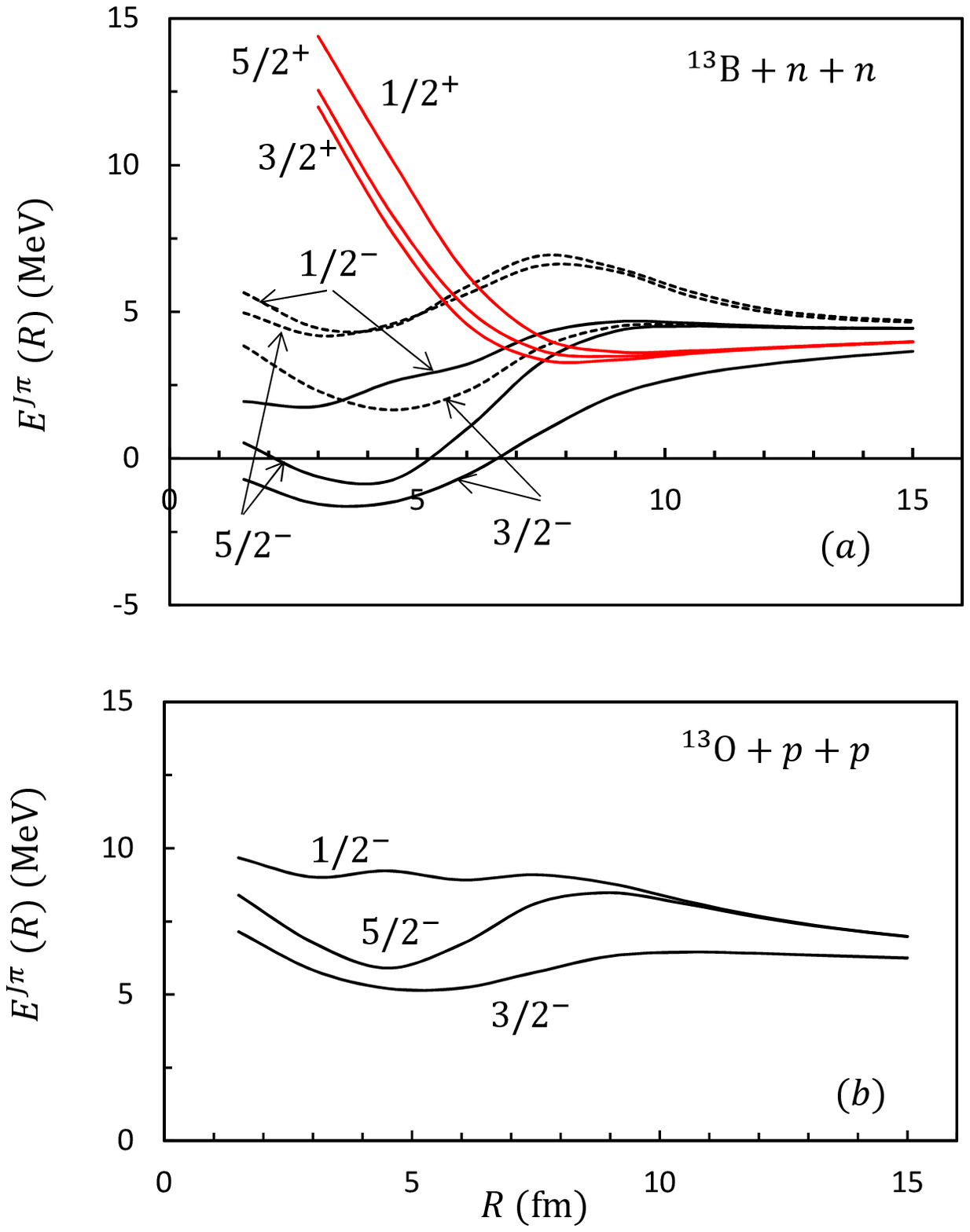,width=8.6cm}
		\caption{(a) $\bnn$ energy curves (\ref{eq19}) for different $J\pi$-values. For negative parity,
			two eigenvalues are displayed. (b) $\oopp$ energy curves.}
		\label{fig_b15_ce}
	\end{center}
\end{figure}

\begin{figure}[htb]
	\begin{center}
 		\epsfig{file=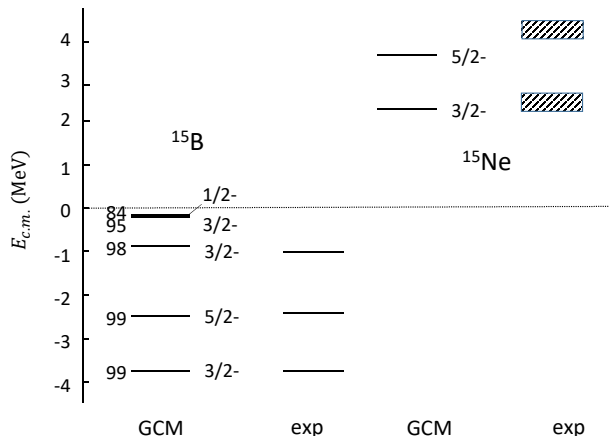,width=8.6cm}
		\caption{$\bo$ and $\neonq$ energy spectra, compared with experiment \cite{KEB05,WMA14}. Numbers at
			 the left of the GCM states correspond to the amount of
			$^{13}$B(g.s.)+n+n configuration (in \%). For $\neonq$, the hatched areas indicate broad resonances. }
		\label{fig_b15_spec}
	\end{center}
\end{figure}

In Fig.\ \ref{fig_b15_ce}, I present the energy curves of $\bo$.  As expected, the lowest energy is 
obtained for $J=3/2^-$.  I may also expect other bound states, corresponding to minima in the energy curves.  
The GCM spectrum, including all generator coordinates, is presented in Fig.\ \ref{fig_b15_spec}.  
The theoretical spectrum shown inf Fig.\ \ref{fig_b15_spec} is remarkably supported by 
experiment \cite{KEB05}, although there is no spin assignment.  The amount of $^{13}$B(g.s.)+n+n
component is shown for the GCM calculation. As suggested by Fig.\ \ref{fig_b15_ce}, the role of
core excitation is small in $\bo$. 
As for $\be$, the positive-parity energy curves are not completely repulsive, but do not support 
narrow resonances.  A more reliable study of positive-parity states in $\bo$
would require introducing $sd$-shell components in the $^{13}$B wave functions.  
This would considerably increase the computer times, and is not feasible at the moment.

The proton and matter radii, given in Table \ref{table2} are in reasonable agreement with experiment.  
The theoretical radius of the $^{13}$B core is 2.31 fm, which means an increase of 0.33 fm for $\bo$. 

Let me briefly discuss the $\neonq$ mirror nucleus, which has been addressed experimentally in
Ref.\ \cite{WMA14} and theoretically in Ref.\ \cite{GXG16}. The ground state and first excited
state have been observed in two-neutron knockout reactions from a $\neon$ beam at 2.5 and 4.4 MeV above
the $\oopp$ threshold. Both states are expected to be broad. 

In the present model, the $\bo$ and $\neonq$
mirror nuclei are studied in the same conditions. The energy curves are shown in Fig.\ \ref{fig_b15_ce}(b).
The $3/2^-$ and $5/2^-$ curves present minima which are associated with the ground and first excited
states. These minima are close to the Coulomb barrier, and clearly suggest continuum
states. According to the higher centrifugal barrier, however, the $5/2^-$ excited state could be
narrower than the $3/2^-$ ground state. The $\neonq$ spectrum is shown in Fig.\ \ref{fig_b15_spec}. Although the GCM energies are
obtained in the bound-state approximation, the results are close to the energies observed experimentally
\cite{WMA14}.

\subsection{$\n$ as $\nnn$ and $\neon$ as $\opp$ systems}
\begin{figure}[htb]
	\begin{center}
		\epsfig{file=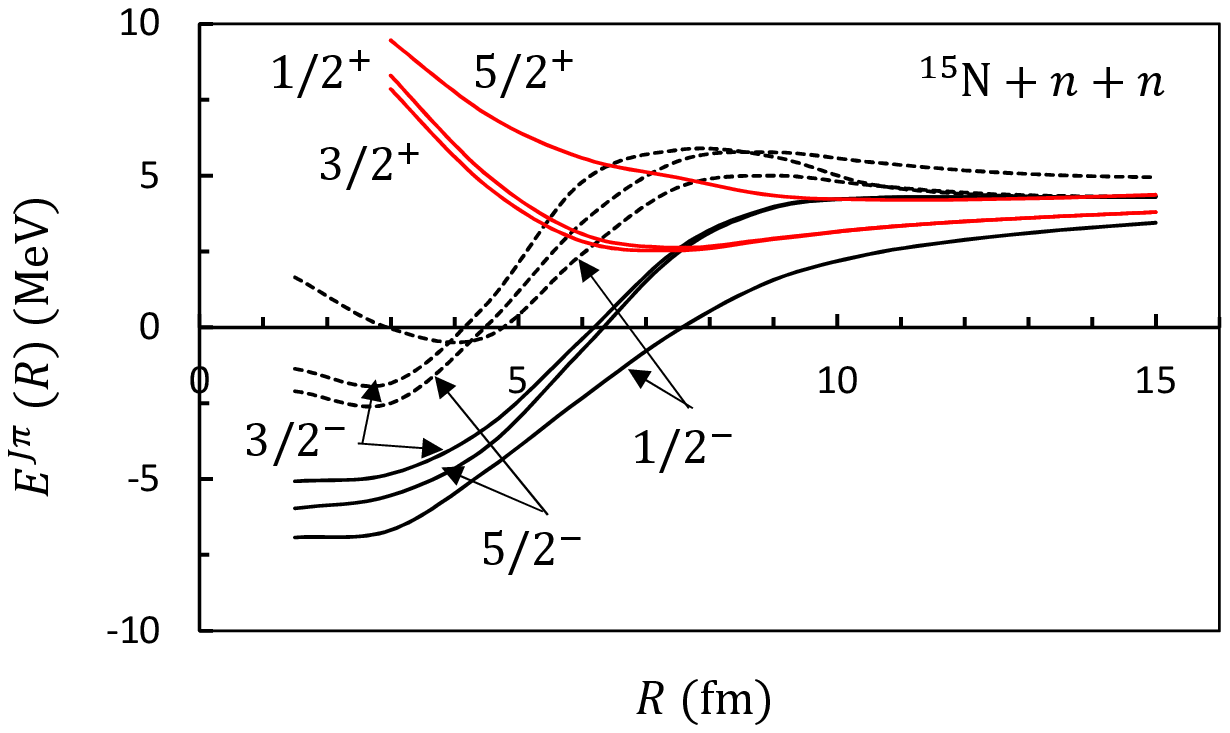,width=8.6cm}
		\caption{$\nnn$ energy curves (\ref{eq19}) for different $J\pi$-values.}
		\label{fig_n17_ce}
	\end{center}
\end{figure}

Microscopic calculations involving a $^{15}$N or a $^{15}$O core are relatively simple since only the ground 
state $1/2^-$ and the first excited state $3/2^-$ are present in the $0\hbar\omega$ shell model.  A previous 
three-cluster microscopic calculation was performed in Ref.\ \cite{TDB96}, where the main goal was 
to address the possible existence of a proton halo in $\neon$.  Figure \ref{fig_n17_ce} shows the energy curves, 
which present a minimum near $R=0$ for negative parity.  This suggests a compact structure for 
negative-parity states, in contrast with the other systems considered here.  

\begin{figure}[htb]
	\begin{center}
		\epsfig{file=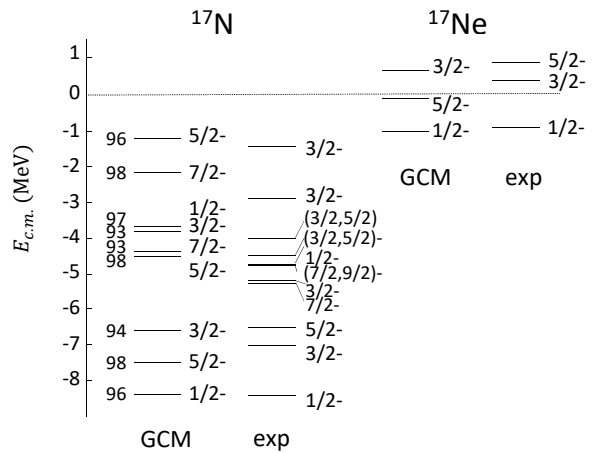,width=8.6cm}
		\caption{$\n$ and $\neon$ energy spectra, compared with experiment \cite{TWC93}. Only negative-parity
		states are shown. Numbers at the left of GCM states correspond to the amount of
		$^{15}$N(g.s.)+n+n configuration (in \%).}
		\label{fig_n17_spec}
	\end{center}
\end{figure}

The $\n$ spectrum is displayed in Fig.\ \ref{fig_n17_spec}.  The general agreement with experiment 
is reasonable but, as in Ref.\ \cite{TDB96} the ordering of the two first excited states is incorrect.
For all states the  $^{15}$N(g.s.)+n+n component is dominant. 
No positive parity state is found, as suggested by the repulsive energy curves of 
Fig.\ \ref{fig_n17_ce}.

The $\neon$ spectrum is studied with the same nucleon-nucleon interaction; the Majorana parameter
is the same as in $\n$. The GCM reproduces the binding energy very well ($-1.05$ MeV in the model,
to be compared to the experimental value $-0.93$ MeV), which shows that the Coulomb shift is
accurately described. 

Electromagnetic transitions have been suggested
to be a valuable tool to investigate the structure of $\neon$ \cite{GPZ05,Fo18}. The E2 transition
probabilities have been studied by relativistic Coulomb excitation with a $\neon$ radioactive beam
\cite{CBF97,CTT02,MWA16}. The $B(E2)$ values, computed without any effective charge, are presented in Table
\ref{tab_be2}. The $B(E2,1/2^- \rightarrow 5/2^-)$ GCM value is in excellent agreement with the latest data of 
Marganiec {\sl et al.} \cite{MWA16}. The experimental value of Chromik {\sl et al.} \cite{CTT02} is larger,
but is probably influenced by nuclear effects which have been dismissed in the analysis \cite{MWA16}. The
transition to the $3/2^-$ state is also well reproduced, which shows that the GCM wave functions are
reliable.

\begin{table}[htb]
	\caption{E2 electromagnetic transition probabilities (in $e^2$.fm$^4$) in $\neon$. No effective
		charge is used.
		\label{tab_be2}}
	\begin{ruledtabular}
		\begin{tabular}{ccc}
			& GCM & Exp.  \\  
			\hline
$1/2^- \rightarrow 5/2^-$ & 92.9 & $90\pm 18$ \cite{MWA16}, $124\pm 18$ \cite{CTT02} \\
$1/2^- \rightarrow 3/2^-$ & 68.0 & $68^{+19}_{-25}$ \cite{CBF97} \\
$5/2^- \rightarrow 3/2^-$ & 7.1 &  \\
		\end{tabular}
	\end{ruledtabular}
\end{table}

The matter radius of the $\n$ ground state, given in Table \ref{table2}, is an excellent agreement 
with experiment \cite{OKS94}.  For $\neon$, however, even if the binding energy is lower, the GCM does 
not support the large difference with $\n$.  I rather confirm the conclusion of Ref.\ \cite{TDB96}, 
that there is no evidence for a proton halo in $\neon$.

\subsection{$\li$ as a $\linn$ system}

The $\li$ nucleus has been studied in many experimental and theoretical works.  The very low 
binding energy ($S_{2n}=0.367$ MeV \cite{AKM17}) is responsible for a remarkable halo structure, 
as suggested by the large r.m.s.\ radius.  The present calculation is made difficult because 
of the large number of Slater determinants (90) involved in the $^{9}$Li core. A previous microscopic
three-cluster study was performed in Ref.\ \cite{De97b}, but with a frozen triangular geometry.
In Ref.\ \cite{VSL02}, the authors describe the core in an $\alpha+t+n+n$ multicluster configuration.

\begin{figure}[htb]
	\begin{center}
		\epsfig{file=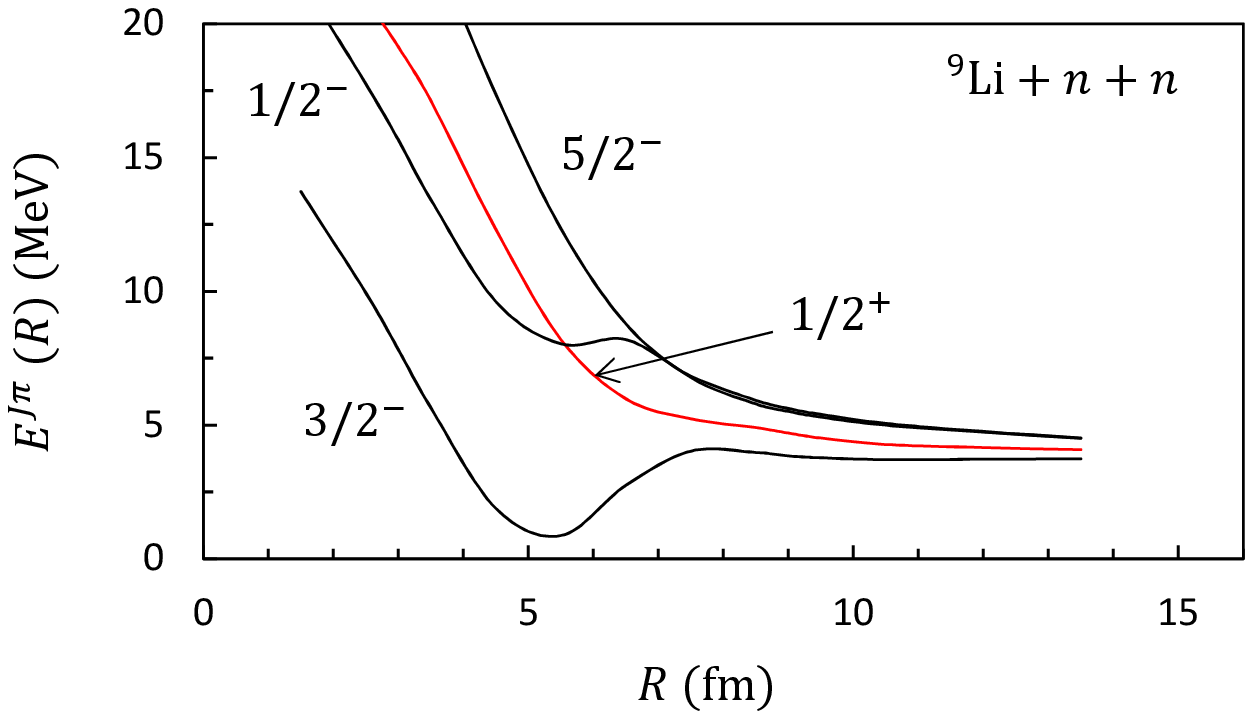,width=8.6cm}
		\caption{$\linn$ energy curves (\ref{eq19}) for different $J\pi$-values.}
		\label{fig_li11_ce}
	\end{center}
\end{figure}

The $\linn$ energy curves are displayed in Fig.\ \ref{fig_li11_ce}. A minimum is obtained for
$J=3/2^-$ and, to a lesser extend, for $J=1/2^-$. The $3/2^-$ minimum corresponds to the
$\li$ ground state. The role of core excitations is illustrated in Fig. \ref{fig_kmax}. When core
excitations are neglected, the state is unbound. The importance of core excitations was already
pointed out in Ref. \cite{VSL02}. In contrast with the previous examples, where the neutron number of the core 
$N=8$ corresponds to a closed shell, the various $^9$Li+n+n configurations are not orthogonal
to each other. Consequently, a ground-state component in $\li$ cannot be estimated.	
 
The proton, neutron and matter radii are presented in Table \ref{table2}. The calculation confirms the large
enhancement of the matter radius, compared to the $^9$Li core (the shell-model value is
$\sqrt{<r^2>_m}(^9{\rm Li})=b\sqrt{17/9}=2.20$ fm). The GCM value is, however, slightly smaller than experiment.
For the proton radius, the experimental $\li$ value (2.467 fm) is significantly larger than the $^9$Li value
(2.217 fm), which suggests that the neutron halo of $\li$ affects the core \cite{SNE06,PMP06}. In the present
model, the $^9$Li proton radius is $\sqrt{<r^2>_p}(^9{\rm Li})=b\sqrt{4/3}=1.84$ fm. The difference between
$\li$ and $^9$Li is therefore 0.13 fm, which is smaller than experiment (0.25 fm) (see a detailed
discussion in Refs.\ \cite{SNE06,PMP06}).

\section{Conclusion}
\label{sec4}
The main goal of the present work is to extend the hyperspherical formalism to microscopic three-cluster
models. The hyperradial functions are expanded over a Gaussian basis using the Generator Coordinate method. 
Then the wave functions are expressed in terms of projected Slater determinants. This approach is well 
adapted to multicluster systems. With a single generator coordinate, it provides an accurate
description of the wave functions, even at long distances. The calculations of the matrix elements, however,
require very long computer times, owing to the seven-dimension integrals necessary for the angular
momentum projection. The extension to $p$-shell cores raises additional difficulties due to $(i)$ the
quadruple sums involved in the matrix elements of two-body interactions; $(ii)$ the presence of several
Slater determinants; $(iii)$ the introduction of core excited states. The calculations are made possible
thanks to an efficient parallelization of the computer code.

The model has been applied to some exotic light nuclei: $\be$, $\bo$, $\neon$ and $\li$. In all cases,
the only parameter (the admixture parameter $M$ involved in the Volkov nucleon-nucleon interaction)
is adjusted on the ground-state energy. For $\be$, the $2^+$ excitation energy is well reproduced,
in contrast with non-microscopic models. The GCM spectrum of $\bo$ is in nice agreement with
the experimental energies. An exploratory study of the $\neonq$ mirror system, which is unbound, is
consistent with the experimental energies, and suggests that the first excited state is narrower than the
ground state. The $\n$ spectrum presents many states which, in general, are fairly well reproduced by the
GCM. The ground state is confirmed to be $1/2^-$, but the order of the first two excited states in
incorrect, as in Ref.\ \cite{TDB96}. In the $\neon$ mirror nucleus, the ground-state energy, as well as E2
transition probabilities are in good agreement with experiment. For $\li$, calculations
are extremely long, owing to the 90 Slater determinants involved in the core. The neutron and matter radii are
explained fairly well, but the proton radius is smaller than the experimental value.

The present model could be extended to deal with three-body continuum states \cite{DD09}. On the other hand, the wave functions could be used as an input to determine scattering cross sections, as it was done for $^6$He for example \cite{De16b}.
	
\section*{Acknowledgments}
P. D. is Directeur de Recherches of F.R.S.-FNRS, Belgium. 
This work was supported by the Fonds de la Recherche Scientifique - FNRS under Grant Number 4.45.10.08.
It benefited from computational resources made available on the Tier-1 supercomputer of the 
F\'ed\'eration Wallonie-Bruxelles, infrastructure funded by the Walloon Region under the grant agreement No. 1117545. 

\appendix
\section{Core wave functions}
\label{appendix}
In this Appendix I give some detail about the core wave functions (\ref{eq4}). The quantum numbers
($I,L,S,T$) are given in Table \ref{table1}. First, the list of functions $\bar{\Phi}_i$ is determined,
and is then used to diagonalize the operators $\pmb{I}^2,\pmb{L}^2,\pmb{S}^2$ and $\pmb{T}^2$. The diagonalization
provides eigenvalues and coefficients $d^{I\nu LST}_i$, which do not depend on the Hamiltonian.

\begin{table}[htb]
	\caption{Quantum numbers ($I,L,S,T$) of the core wave functions. $n$ is the degeneracy.
		\label{table1}}
	\begin{ruledtabular}
		\begin{tabular}{lccccc}
			& $I$ & $T$ & $S$ & $L$ & $n$  \\  
			\hline
			$^9$Li &	1/2 & 3/2 & 1/2 & 0 & 1 \\
			&	  &   &   & 1 & 2 \\
			&	  &   & 3/2 & 1 & 1 \\
			&	  &   &   & 2 & 2 \\
			&	  &5/2 & 1/2 & 1 & 1 \\
			&3/2 &3/2& 1/2 & 1 & 2 \\
			&  & &   & 2 & 2 \\
			&  & & 3/2 & 0 & 1 \\
			&  & &   & 1 & 1 \\
			&  & &   & 2 & 1 \\
			&  &5/2& 1/2 & 1 & 1 \\
			&5/2 &3/2& 1/2 & 2 & 2 \\
			&  & &   & 3 & 1 \\
			&  & & 3/2 & 1 & 1 \\
			&  & &  & 2 & 1 \\
			&7/2 &3/2& 1/2 & 3 & 1 \\
			&  & & 3/2 & 2 & 1 \\
			& & & & & \\
			$^{12}$Be &0 & 2 & 0 & 0 & 1 \\
			&  &   & 1 & 1 & 1 \\
			&1 & 2 & 1 & 1 & 1 \\
			&2 & 2 & 0 & 2 & 1 \\
			&  &   & 1 & 2 & 1 \\
			& & & & & \\
			$^{13}$B &1/2 & 3/2 & 1/2 & 1 & 1 \\
			&3/2 & 3/2 & 1/2 & 1 & 1 \\
			&  &   & 1/2 & 2 & 1 \\
			&  &   & 3/2 & 0 & 1 \\
			&5/2 &3/2 & 1/2 & 2 & 1 \\
			& & & & & \\
			$^{15}$N &1/2 & 1/2 & 1/2 & 1 & 1 \\
			&3/2 & 1/2 & 1/2 & 1 & 1 \\
		\end{tabular}
	\end{ruledtabular}
\end{table}

\clearpage
%

\end{document}